# Non-contact, non-destructive mapping of thermal diffusivity and surface acoustic wave speed using transient grating spectroscopy

Abdallah Reza[1†], Cody A. Dennett[2,3], Michael P. Short[3], John Waite[4], Yevhen Zayachuk[4,5], Christopher M. Magazzeni[4], Simon Hills[1], Felix Hofmann[1*]

[1]Department of Engineering Science, University of Oxford, Parks Road, Oxford, OX1 3PJ, UK

[2] Materials Science and Engineering Department, Idaho National Laboratory, Idaho Falls, ID, 83415, USA

[3]Department of Nuclear Science and Engineering, Massachusetts Institute of Technology, Cambridge, Massachusetts 02139, USA

[4]Department of Materials, University of Oxford, Parks Road, Oxford, OX1 3PH, UK

[5]UK Atomic Energy Authority, Culham Science Centre, Abingdon, OX14 3DB, UK

[†] *mohamed.reza@eng.ox.ac.uk*

[*] *felix.hofmann@eng.ox.ac.uk*

## Abstract

We present new developments of the laser-induced transient grating spectroscopy (TGS) technique that enable the measurement of large area 2D maps of thermal diffusivity and surface acoustic wave speed. Additional capabilities include targeted measurements and the ability to accommodate samples with increased surface roughness. These new capabilities are demonstrated by recording large TGS maps of deuterium implanted tungsten, linear friction welded aerospace alloys and high entropy alloys with a range of grain sizes. The results illustrate the ability to view grain microstructure in elastically anisotropic samples, and to detect anomalies in samples, for example due to irradiation and previous measurements. They also point to the possibility of using TGS to quantify grain size at the surface of polycrystalline materials.

## Introduction

Advances in the characterisation of engineering material properties are essential to keep up the pace of material innovation. For applications in heat engines, gas turbines, and extreme environments, such as nuclear fusion reactors, mechanical and physical properties, as well as their evolution in service, are very important [1], [2]. Transient grating spectroscopy (TGS) presents a multi-modal non-destructive characterisation method that enables rapid insight into both the elastic and thermal transport properties of materials [3]–[7].

TGS measures the surface acoustic wave (SAW) speed, from which the elastic constants can be obtained [8]. Not only does the SAW speed provide quantitative insight into the elastic properties, it can also be used to pin-point microstructural changes, such as void swelling in irradiated materials [9] and for the evaluation of the properties of micron-thick surface layers [10]–[12]. Thermal diffusivity measurements from TGS have been used to detect the presence of crystal defects and impurities in metals [13], as well as to monitor irradiation-induced microstructural evolution for example in single-crystal niobium [14] and tungsten [13], [15], [16].

TGS presents several key advantages over other more conventional thermal diffusivity and acoustic property measurement methods: It is non-contact, non-destructive, and quite rapid (< 2 s for a spot measurement). The only requirement for the measurement is a flat sample surface polished to a mirror finish. Heterodyne detection [17] and the newer dual-heterodyne detection [7] allow high sensitivity measurements, thereby reducing the laser power required and thus the risk of surface modification due to the laser exposure. TGS also allows the depth-sensitive probing of properties, making it possible to specifically probe micron thick surface layers of interest, for example in semiconductors [6], or in ion-irradiated nuclear materials [13]. The nature of the measurement also does not impose an intrinsic limit to the sample temperatures at which TGS can be performed, making TGS attractive for in-situ measurements, e.g. during sample heating.

A substantial drawback of TGS measurements has been the need for frequent manual intervention when carrying out multiple measurements, as the geometry is rather sensitive to even small sample misalignments. This has prevented the automated measurement of large spatial maps, rather only point measurements have been possible. Setting these up, especially when capturing of spatial variation is required, was tedious and time consuming. Spatial mapping of SAW velocity tends to be less challenging than the mapping of thermal diffusivity, since it only requires the frequency component in the signal rather than the entire thermal decay. Rapid spatial mapping of SAW velocity using TGS excitation with a point probe has been successfully implemented, however with significantly larger probing depths ( > 10 μm) [18], [19]. Thermal diffusivity mapping using time-domain thermo-reflectance (TDTR), with finer spatial resolution (~2 μm), has also been implemented [20]. The advantage of TGS over these approaches is that it provides both the SAW velocity and thermal diffusivity in one measurement and with high temporal resolution (< 2s for a spot measurement).

Here we present new developments that make TGS measurements much more robust to small sample misalignments. This makes it possible to accommodate samples with increased surface roughness, and enables the automated thermal diffusivity and SAW velocity mapping of large sample areas.

## THE TRANSIENT GRATING SPECTROSCOPY SETUP

In the following we provide a brief summary of the TGS method. A detailed description of TGS theory and methodology, can be found elsewhere [3]–[7].

In TGS, two coherent laser pulses are crossed with a well-defined angle at the sample surface. The crossing of the beams creates a spatially periodic interference pattern on the sample surface with a tunable wavelength λ. Absorption of the light from this intensity grating results in a spatially periodic heating of the sample, i.e. the formation of a temperature grating. Thermal expansion leads to a displacement of the sample surface, creating a displacement grating. Together the temperature and displacement gratings are referred to as the "transient grating". A third "probe beam" (continuous wave) incident on the excited sample surface is diffracted by the transient grating, and reflected by the sample surface. As heat in the temperature grating diffuses from peaks to troughs and into the bulk, the grating decays. The rate of decay is linked to the thermal diffusivity of the sample [4], and manifests itself in the intensity decay of the diffracted probe signal as seen in Fig. 1.

Rapid thermal expansion of the sample also launches two counter-propagating surface acoustic waves (SAWs), with the same well-defined wavelength, λ, as the transient grating. The SAW's superimpose an oscillation on the transient grating in the sample that results in a superimposed oscillation in the TGS signal, as seen in Fig. 1. The SAW frequency depends on the elastic properties of the material [21]. To extract SAW frequency and thermal diffusivity, the TGS traces are fitted with the expressions given in [4]. There are a number of examples for this [13], [15], [16], [22] with custom routines in MATLAB frequently being used to do the fitting.

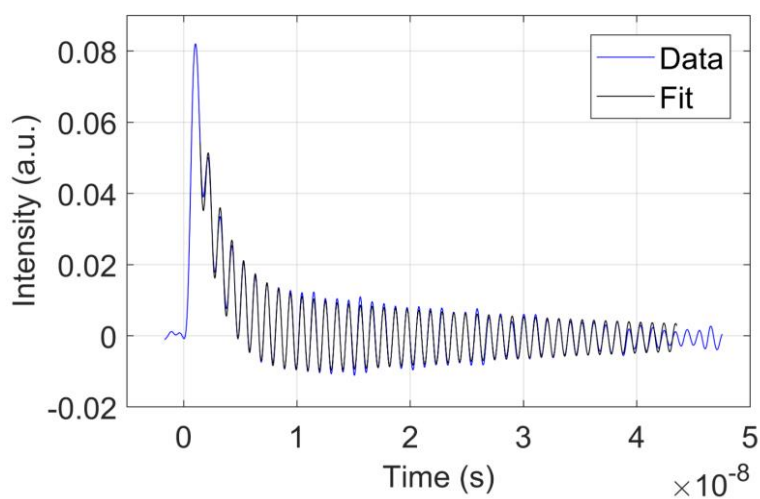

*Figure 1:* Sample TGS trace from an annealed, polycrystalline, 99.99% purity tungsten sample. Also shown is a fit to the experimental data using the expression outlined in [13].

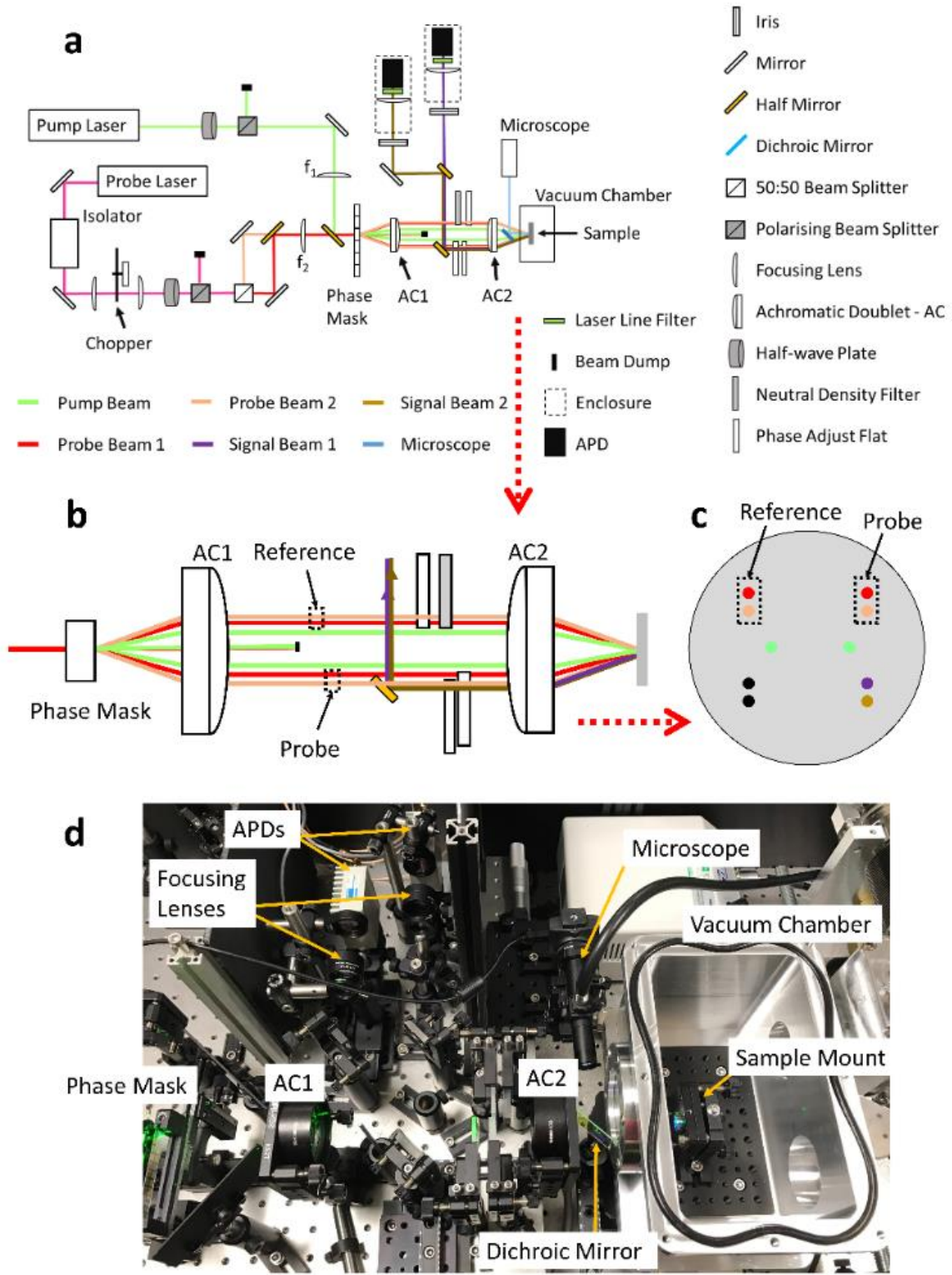

*Figure 2:* Layout of the TGS setup. (a) System overview indicating key components. (b) Layout of the 4f imaging system with the two achromats labelled AC1 and AC2. For clarity the dichroic mirror is not shown in (b). (c) Cross-sectional layout of the optical beams at achromat AC2, looking in the direction of the incident pump beams. Shown are the pump (green), probe (red and cream), signal (purple and brown) and the reflected "waste" probe beams (black). (d) Image of the setup from the phase mask onwards with key components labelled.

The following subsections detail the various parts of the setup and improvements implemented here.

## Beam Generation

The pump beams are obtained from a 532 nm Q-switched microchip laser (Teem Photonics, PNG-M02010-120), with a pulse length of ~0.5 ns, maximum energy per pulse of 20 μJ and 1 kHz repetition rate. A combination of a half wave plate and a polarising beam splitter (PBS) are used to control the beam power, giving ~2 μJ per pulse at the sample. An advantage of this system is that the laser power can be regulated without having to interfere with the laser or beam path. An additional advantage is that the optics downstream do not have to accommodate excess laser beam energy, which is re-directed to a beam dump by the PBS.

The TGS setup developed in this study is based on previous iterations and includes heterodyne detection [5], the box-car geometry [23], and the more recently developed dual-heterodyne TGS configuration [7]. Heterodyne detection increases the signal amplitude, hence reducing the required laser beam power. The box-car geometry uses common optics for all pump and probe beams, as shown in Fig. 2(b), and permits very high relative phase stability between beams [23]. Dual-heterodyne detection allows the simultaneous detection of two heterodyne phases without the need for manual intervention. This improves the time resolution and repeatability.

The pump beam (green in Fig. 2 (a)) is focussed (using a plano-convex spherical lens with a focal length of 150 mm, lens $f_1$ in Fig. 2(a)) onto a phase mask with a spot size of 140 μm ($1/e^2$). The phase mask is optimised such that ~50% of the incident light is directed in the +1 and -1 diffraction orders. All other diffracted orders, as well as the $0^{th}$ order beam are blocked. The transmitted first order diffracted beams are re-imaged onto the sample using two achromatic doublets (AC1 and AC2) in a 4f imaging system (the box-car) with the pump beams maintained in the horizontal plane. Use of achromatic doublets minimises chromatic aberrations due to differences in the probe and pump beam wavelengths. It should be noted that it is essential to have different probe and pump wavelengths in order to be able to spectrally separate the TGS signal from pump beam scatter at the detectors.

The probe beams are generated using a 561 nm Coherent Genesis MX 561 laser with large coherence length (500 mW max. power, single-longitudinal mode). The continuous wave beam is chopped at 1 kHz using a mechanical chopper with a 25% duty cycle, and synchronised to the output of the pump laser. In a similar fashion to the pump beams, the probe beam power is regulated using a half wave plate and PBS combination. Back-reflected beams are prevented from re-entering the laser using a Thorlabs polarisation-dependent isolator (IO-5-560-HP, 560 nm). Since the dual heterodyne technique requires two sets of probe beams, a 50:50 beam splitter is used to split the probe beams evenly (red and light brown beams in Fig. 2(a)). Separate mirrors are used to align these probe beams parallel to the optical axis of an aspheric focusing lens (f= 100 mm, lens $f_2$ in Fig. 2(a)), but at different heights. This improvement from previous TGS iterations allows independent adjustment of the probe beams, greatly simplifying alignment. The lens $f_2$ is then used to focus the probes onto the phase mask with a 90 μm spot ($1/e^2$). Similar to the pump beams, only the first order diffracted beams from the phase mask are transmitted, whilst all other diffraction orders and the direct beam are blocked. For each probe the +1 diffracted beam acts as the "probing arm" and the -1 diffracted beam as the "reference arm", as shown in Fig. 2(b) and (c). The reference beams are attenuated by a factor of 100 using an ND2 filter. Phase adjust flats are used on the probing arm to allow accurate shifting of the probe beams' phase with respect to their respective reference beams, with relative phases $\varphi_1$ and $\varphi_2$ for the two probes. An additional optical flat is introduced to the upper reference beam, to match the total path length of the reference and probe beams and achieve a higher phase stability. A cross-sectional view of the pump, probe, and signal beams at achromatic lens AC2, prior to focusing on the sample, is shown in Fig. 2(c).

## Sample Environment

In the reflection TGS geometry, SAW measurements can be conducted at atmospheric pressure. However, thermal diffusivity measurements require a vacuum environment to remove the effects of excitations in air near the sample surface commonly observed in ambient conditions [5]. For this purpose, a vacuum chamber was built (see Fig. 2(d)). An Edwards Vacuum T-Station 85 setup that utilises a turbo-molecular pump and dry scroll pump is used to evacuate the chamber. A vacuum level of $10^{-3}$ mbar is sufficient for thermal diffusivity measurements at room temperature. Calibration tungsten samples returned a thermal diffusivity of ~$(6.6 \pm 0.2) \times 10^{-5}$ $m^2s^{-1}$ at this vacuum level, which agrees very well with previous measurements, as well as the values given in the literature [24]–[26].

To create the TGS maps, automated sample stages are used in the directions perpendicular to the axis of the beams (Thorlabs motorised lab jack MLJ150 and Physik Instrumente L-511 linear stage for the vertical and horizontal translation, respectively). Translation along the beam is carried out using a manual stage. The range of motion currently available is 100 mm and 50 mm in the horizontal and vertical directions respectively. At present the size of the largest thermal diffusivity scan is limited by the size of the viewport on the vacuum chamber, which is ~50 mm in diameter.

## Targeted Measurements

Targeting TGS measurement to a particular location on the sample surface is important for the investigation of specific microstructural or sample features. Here we incorporate an Edmund Optics in-line microscope with 3X magnification into the setup (TECHSPEC® In-Line CompactTL Telecentric Lens). At a fixed working distance of 110 mm this microscope has a field of view of 1.6 x 1.1 $mm^2$ when used with a Thorlabs DCC1645C 1.3 megapixel camera. The pixel size in this case is 1.3 µm, and features as small as 10 µm can be readily resolved. The beam geometry and sample environment only permits viewing of the sample through the same viewport also used by the laser beams. A dichroic mirror was used to allow the passage of the microscope illumination without affecting the TGS beams. The dichroic (Thorlabs long-pass dichroic with a 505 nm cut-off), was chosen such that it transmits the 532 nm pump and 561 nm probe beams and but reflects the blue microscope illuminator light. This enables viewing of the sample in-situ, during TGS measurements using the microscope. The microscope and dichroic mirror setup are shown in Fig. 2(a) and (d). The presence of the dichroic mirror in the path of the pump beams does not affect the wavelength of the TGS grating on the sample since it acts on both arms of the pump and probe beams. This was confirmed by comparing the SAW frequency in a calibration tungsten sample, with (468.7 ± 0.5 MHz) and without (468.8 ± 0.4 MHz) the dichroic mirror in place.

### Detection

The duration of the TGS signal that follows the 0.5 ns pump pulse may range from 50 ns (for small grating periods and samples with high thermal diffusivity e.g. tungsten) up to several microseconds (for large grating periods and samples with low thermal diffusivity e.g. ceramics). In this TGS implementation, silicon avalanche photo diodes (APDs) were used (Hamamatsu C5658). The lower and upper cut-off frequencies of these APDs are 50 kHz and 1 GHz respectively. The upper cut-off is evident in Fig. 3(b) where the SAW frequencies measured are accurate up to ~1.4 GHz. A Teledyne LeCroy WavePro 760 Zi, 6 GHz digital oscilloscope is used to readout the APD signals. Scripts in MATLAB [27] are used to control the oscilloscope data acquisition.

In the TGS setup, depending on the roughness of the sample surface, a significant amount of scattered pump light can accompany the signal beams up to the detectors. Laser line filters, as well as an iris positioned in front of each APD are used to remove this scatter (see Fig. 2). Scatter from other components in the setup is blocked by enclosing the detection optics, including the APDs. In addition, plano-convex lenses are used to focus the signal beams on to the 0.5 mm diameter active region of the APDs. This is discussed in further detail in the section 'Modification for samples with uneven surfaces'.

### Validation Experiments

Validation experiments were carried out to benchmark performance of the setup. This included thermal diffusivity and SAW frequency measurements on tungsten samples while the sample was taken to vacuum from ambient conditions, measurements over long periods of time, as well as varying pump and probe powers.

To explore the upper cut-off frequency of the setup, a silicon wafer (0.32 mm thick, with a <111> out-of-plane orientation), was measured using various nominal grating periods from 2.8 to 8.8 μm on the sample. Fig. 3 shows that up to a frequency of 1.4 GHz there is the expected linear relationship between SAW frequency and $1/\lambda$, where the slope of the line gives the SAW velocity. This suggests an upper frequency limit of the setup of ~1.4 GHz, consistent with the excitation laser pulse length, as well as the APD cut-off frequency. The SAW frequency for a tungsten single crystal (1.0 mm thick, with a <111> out-of-plane orientation) was also measured for the same grating periods and is plotted in Fig. 3.

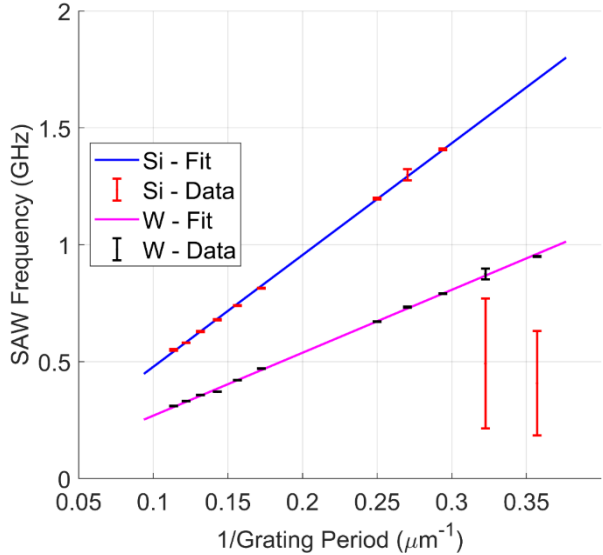

**Figure 3:** Surface acoustic wave frequency vs. 1 / nominal grating period for a <111> silicon single crystal and a <111> tungsten single crystal.

To assess the repeatability of the measurements, and to determine whether the pulsed laser heating affects the TGS measurements, continuous scans were carried out over a period of 5+ hours on a single spot in pristine and self-ion implanted tungsten (0.32 dpa, 20 MeV, RT). The tungsten samples were polycrystalline with random crystalline orientation (tungsten is almost perfectly elastically isotropic [28]). Further details of the samples, the implantation and its effects can be found in [15]. Fig. 4(a) and (b) show that there is no change in the measured value of the thermal diffusivity or the SAW speed over the entire duration of the test for either of the samples. Superimposed on the plots are data from literature for pristine tungsten [28], [29]. The SAW speed literature data [28] is for single crystal tungsten, and represents upper and lower bounds of SAW speed due to the small elastic anisotropy of tungsten. As expected our values lie within this range, since our measurements are averaged over multiple grains with varying orientation. This is due to the measurement spot (~90 μm) being larger than the average grain size (~20 μm) [15]. If laser heating were significant, the measured thermal diffusivity in pristine tungsten would be lower than the room temperature book value. Fig 4(a) does not show this effect. Similarly the lack of long-term evolution of thermal diffusivity and SAW speed suggests that the heating does not modify the underlying defect microstructure in the self-ion-implanted tungsten sample.

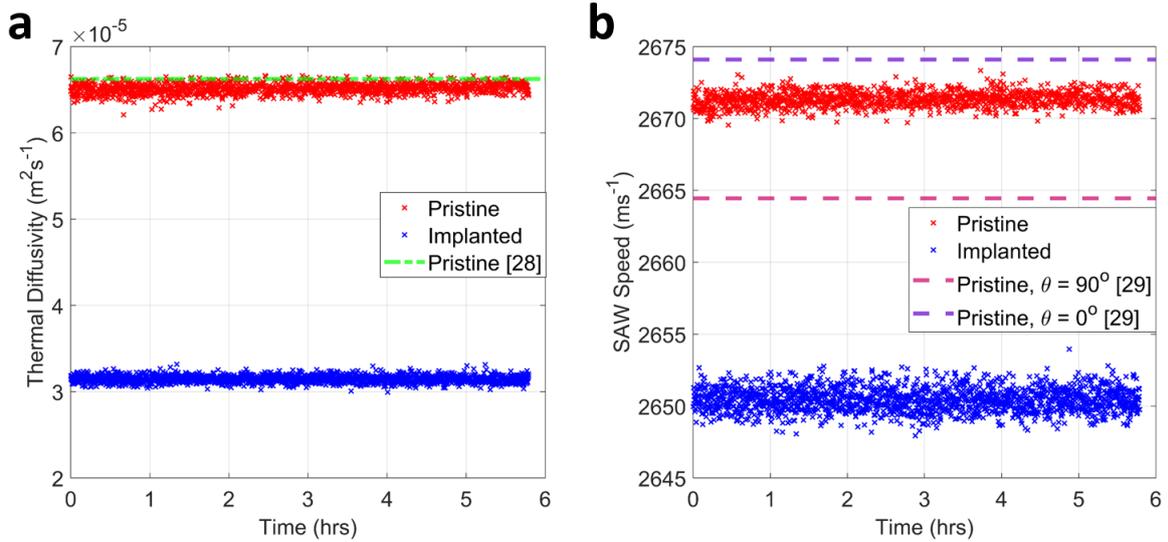

**Figure 4:** Long duration TGS scans on pristine and self-ion implanted tungsten. (a) Thermal diffusivity and (b) SAW speed. Also superimposed on the plots are the book values for pristine tungsten. The angles in (b) denote the SAW propagation direction in a <110> crystal.

The standard deviation of the thermal diffusivity and SAW speed measurements is 1% and 0.03% respectively. This is an indication of the repeatability of the measurements and depends on the number of averages taken [7], [19] . In this study 2000 – 5000 averages were deemed sufficient. The absolute error in the thermal diffusivity, considering the deviation from the book value for the pristine tungsten sample [29], is ~4%. This is an improvement over other thermal diffusivity measurement techniques that report values between 5 and 10 % [30]. The TDTR thermal diffusivity mapping study [20] reported an uncertainty of 8% in the thermal diffusivity. The SAW speed uncertainty (0.03%) is comparable to previous measurements (0.5 - 0.05%) [19].

### Modification for samples with uneven surfaces

Dual heterodyne TGS requires six beams to cross at the same position on the sample surface with a precision greater than 50 µm (see Fig. 2(a, b and c)). The signal beams leaving the sample surface consist of the two diffracted probe beams, each combined with a heterodyne reference beam. Slight changes in sample surface orientation lead to a pointing error and hence misalignment of the signal beams at the detector. Any unevenness of the sample surface will change the exit angle of the signal beams, that then shift laterally downstream of AC2 as illustrated by red arrows in Fig. 5. When the setup is configured for single point measurements, this is not a problem as the sample orientation can be aligned. However, for the automated mapping of larger samples that are not perfectly flat, the exit angle of the signal beams may vary depending on the location on the sample. The consequent lateral shift of the signal beams can be sufficient to lead to a loss of signal on the APDs, which have a 500 μm diameter sensitive area.

To compensate for sample slope errors, plano-convex focusing lenses were placed in the path of the signal beams with the detector active area at the focal plane. With this modification, even with a lateral or slight directional shift of the signal beams, the signal still reaches the detector, enabling the creation of larger TGS maps. To quantify this capability, a flat, well-polished tungsten sample was aligned normal to the beams. The maximum sample tilt angle in the vertical and horizontal plane that still gave a signal with (1/e) of the maximum signal amplitude was then measured. Without the focussing lenses in place the maximum tilt, α, was 0.2° and 0.1° in the horizontal and vertical planes respectively. With the lenses in place, the sample could be tilted by up to 2.0° and 1.5° in the horizontal and vertical planes, respectively. This means that our setup can tolerate an order of magnitude larger sample roughness than in previous works without this modification. It should be noted that a 2° inclination in the sample surface changes the transient grating period by ~0.06%. This is comparable to the repeatability of the SAW speed measurement and substantially less than the uncertainty in the thermal diffusivity measurement.

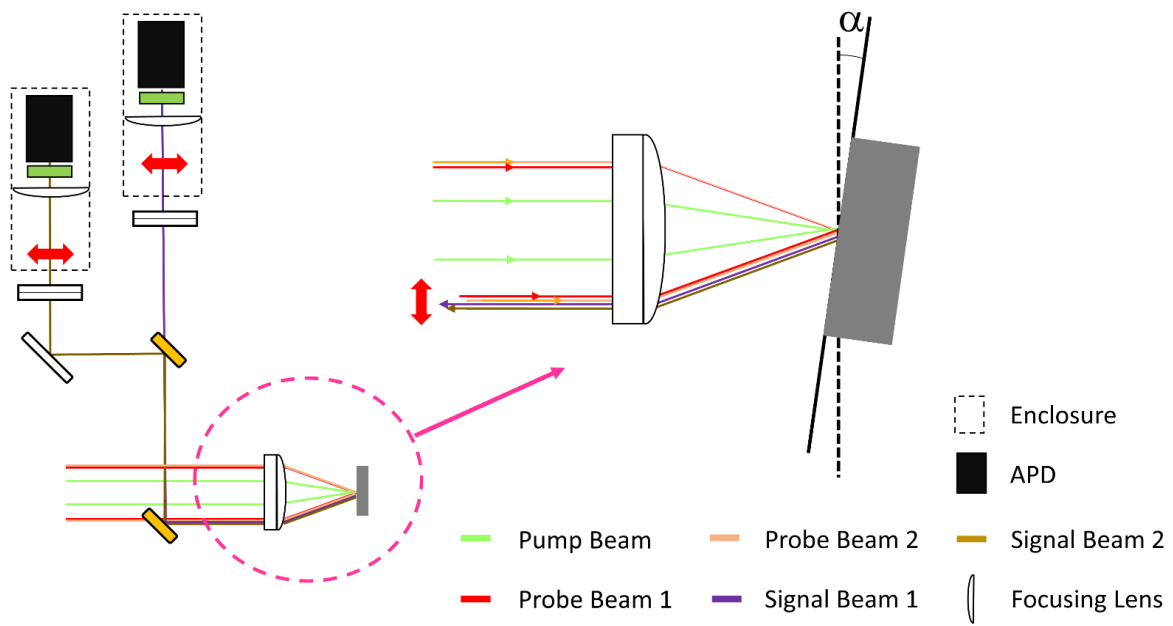

**Figure 5:** Schematic of the sample arrangement for the experiment to determine the maximum permissible sample surface angle. The red double-arrows indicate the lateral shift of the beams due to the sample tilt. The optics key is the same as in Fig. 2 (a).

## Results and Discussion

### Deuterium Exposed Tungsten

The TGS setup was used to study the effects of deuterium ion exposure on tungsten, a promising contender for fusion reactor armour components. Detailed results of this study are provided in [16]. Here the same sample is used to illustrate the 2D mapping capability of the TGS setup.

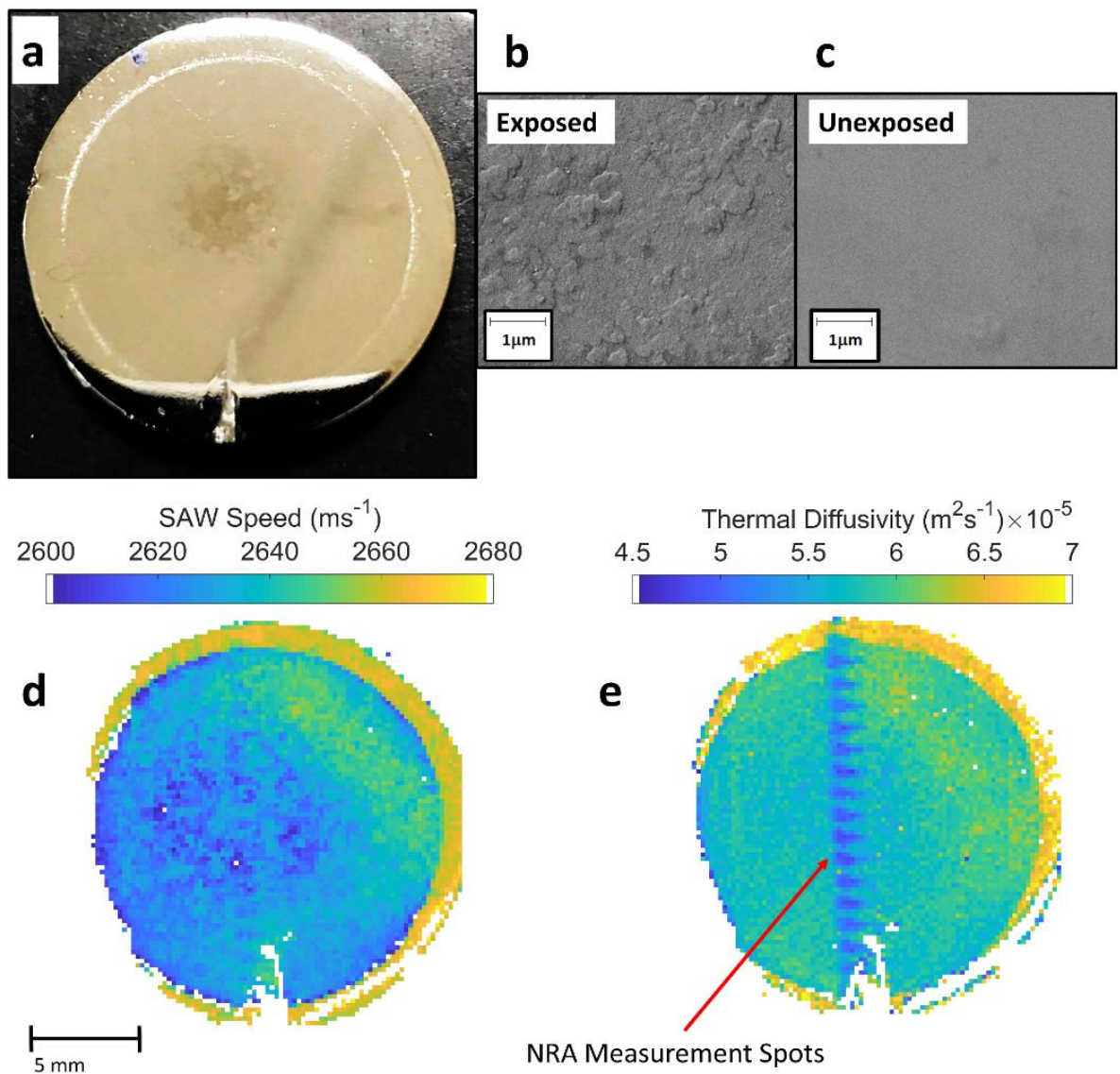

**Figure 6:** Deuterium-exposed tungsten sample. The central region was exposed to deuterium, whilst the edge of the sample was shielded during the exposure. (a) Optical image of the sample. SEM micrographs of the central exposed region (b) and unexposed edge (c). (d) SAW map of the whole sample measured by TGS. (e) Thermal diffusivity map. The same scale bar applies to (a), (d) and (e).

The tungsten sample was exposed to deuterium ion fluences of ~$10^{27}$ ions m$^{-2}$ with Gaussian flux and temperature profiles ($T_{max}$~450K in the centre of the sample). The edges were shielded during exposure. A circular scratch, visible in Fig. 6(a), marks the boundary between exposed and unexposed regions. Deuterium exposure leads to the formation of surface and sub-surface blisters, as seen in Fig. 6(b). A detailed discussion of the responsible mechanisms is provided elsewhere [31]–[33]. Our previous work showed a significant degradation of thermal diffusivity and reduction in SAW speed in the exposed area [16]. Using the mapping capability of the TGS setup we have measured a map of thermal diffusivity and SAW speed that spans the entire sample (Fig. 6(d) and (e)). It consists of over 9000 spatial locations with a step size of 200 μm between measurement points. A grating period of 2.758 ± 0.002 μm on the sample surface was used, such that the measurements will be dominated by the ~1 μm thick deuterium affected surface layer [4].

The SAW speed and thermal diffusivity maps in Fig. 6, clearly show higher values for thermal diffusivity and SAW speed in the pristine unexposed regions. The measured values are in very good agreement with literature values for well-annealed tungsten [24], [26], [28], [34]. Reduced SAW speed and lower thermal diffusivity due to the ion exposure are evident in the central region of the maps.

An interesting feature is a line of points with significantly further reduced thermal diffusivity that crosses the central region of the sample (see Fig. 6(e)). These positions correspond to the locations of previous nuclear reaction analysis (NRA) measurements used to obtain depth profiles of the deuterium concentration in the samples [31]. NRA measurements bombard the sample with particular reactive nuclei and detect the emission of ionising radiation from the resulting nuclear reaction [35]. In this case the NRA technique used helium ions of different energies from 500 keV to 4.5 MeV [31], [36]. These energies are sufficiently high to cause irradiation damage [37]. Given that the bombardment energies are up to a few MeV, the damaged layer is expected to be several microns thick, hence affecting the TGS measurements [38]. Indeed, our previous measurements showed that even small doses of helium injected at these energies can lead to a dramatic reduction in room temperature thermal diffusivity of tungsten [13]. The mapping capability has effectively spotted this additional irradiation damage due to the NRA measurements. By conducting spot measurements or line scans only, these spurious effects of NRA measurements could have easily been missed, highlighting the benefit of the 2D mapping capability of the TGS setup.

Linear friction welded Ti6Al4V

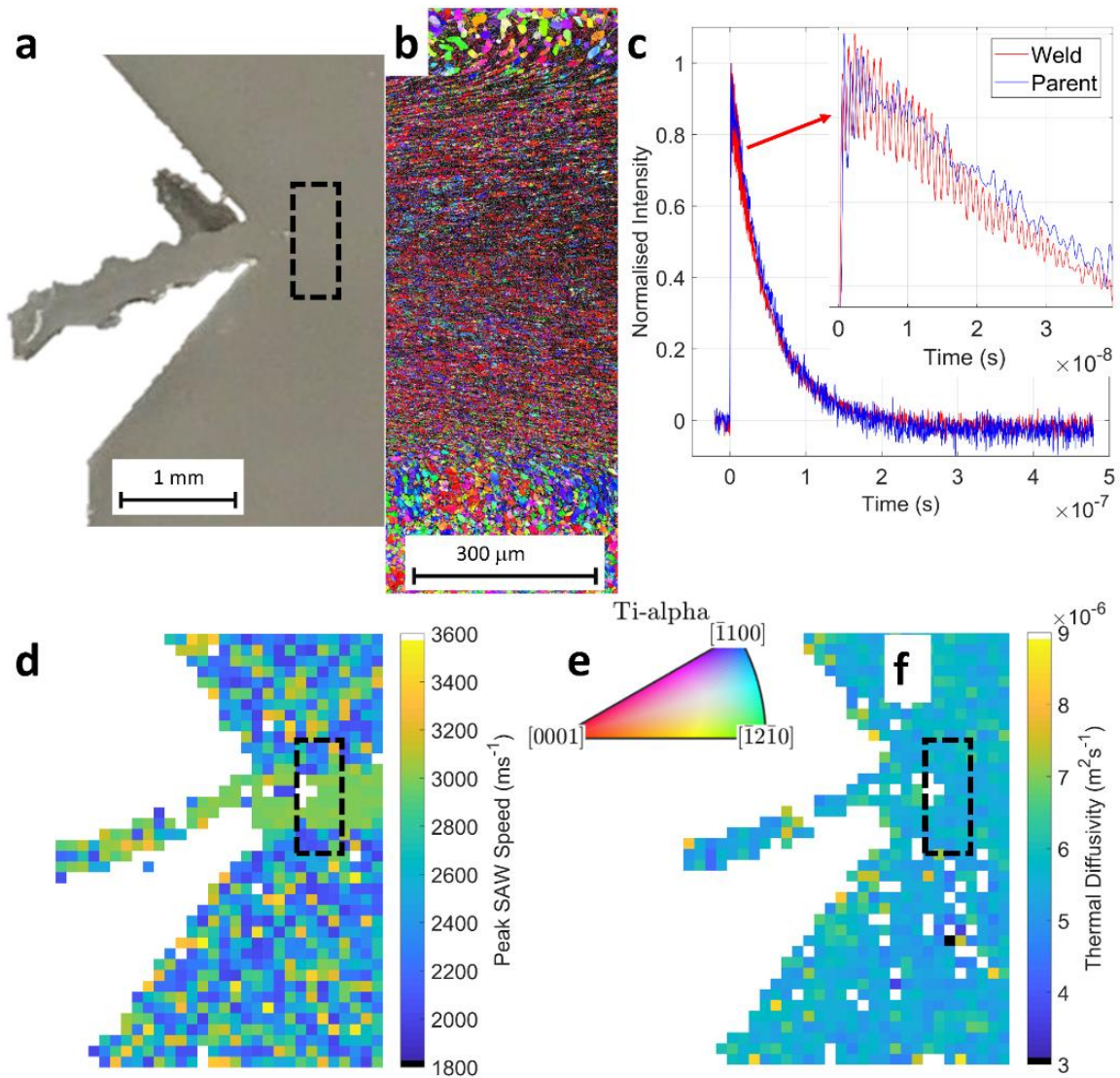

**Figure 7:** Linear friction welded Ti6Al4V sample. (a) Optical micrograph showing the weld and flash extruded from the weld line. (b) EBSD scan showing grain sizes and orientations, (c) sample TGS traces from the weld and parent regions, (d) TGS SAW speed map, (e) EBSD colormap and (f) TGS thermal diffusivity map. The location of the EBSD scan is demarcated by the dashed region. The same scale bar applies to the TGS maps and the optical image.

We also carried out 2D scans of a linear friction welded Ti6Al4V aerospace alloy. Friction welding is increasingly popular due to its advantages over fusion welding, such as avoiding of melting, lower porosity and fewer impurities due to the absence of a filler material [39], [40]. However, there still is a significant change in the microstructure in the weld region that can lead to changes in material properties.

Here we considered a 5x10 mm, 2mm thick section through a Ti6Al4V friction weld. An optical micrograph of the central region of this sample is shown in Fig. 7(a). Phase maps of the parent material from EBSD revealed it to be 99% Ti-alpha and less than 1% Ti-beta. An electron back-scatter diffraction (EBSD) scan of central sample region shows that the weld line has a width of ~500 µm (Fig. 7(b)), with much smaller grains in the weld (~1 µm) than the surrounding parent material (~10-15 µm). EBSD also shows that the material in the weld is strongly textured, whereas grains in the parent material are more randomly oriented. Interestingly even in the parent material there are domains with grains of similar orientation, seen as red and blue regions in Fig. 7(b) reminiscent of macrozones observed in this alloy [41]–[43], which corresponds to crystallographic orientations of [0001] and [$\bar{1}$000]. These macrozones are also identifiable in the SAW map in Fig. 7(d), where there are regions of blue and yellow/green, which correspond to SAW speeds of 2200-2400 ms$^{-1}$ and 2800-3200 ms$^{-1}$. SAW speed in anisotropic materials depends on the crystal orientation and TGS wavevector direction [28]. Hence we expect such variations in the SAW speed given the elastically anisotropic hexagonally close packed (HCP) structure of Ti-alpha. Also noticeable in the map is the uniform SAW speed in the weld. This uniformity in the SAW speed is due to the uniformity of texture in the weld. This demonstrates the ability of the setup to identify regions of uniform orientation and random orientation in elastically anisotropic materials. The SAW speed depends on the elastic modulus alone for elastically isotropic materials and on the stiffness tensor, wave vector direction and crystal orientation for elastically anisotropic materials [8], [28]. Hence this method can be used to quantify the effect of welds on the elastic properties of macroscopic components.

The thermal diffusivity map in Fig. 7(e) shows little variation of thermal diffusivity, indicating that thermal transport in the weld is very similar to that in the parent material. This is a desirable result, which indicates that the weld does not affect the overall thermal conductivity of the component. As is the case with the larger SAW maps, the larger thermal diffusivity map allows us to identify inhomogeneous regions that may lead to anomalies in the dissipation of heat. The fact that we are able to create such maps encompassing the weld, the parent material, as well as the weld flash (material extruded at the joint line during the friction welding process), demonstrates the capabilities of TGS for rapid characterisation of macroscopic engineering components.

Fig. 7(c) shows another interesting feature. The SAW oscillations in the TGS signal persist for longer in the weld than in the parent material. This is believed to be a result of the grain size and orientation. In the weld, the grains are smaller than the TGS wavelength and have similar orientation (strong texture). Hence when the SAW move from one grain to the next, the SAW velocity and wavelength are very similar due to the similar crystal orientation. The probe still meets the diffraction condition for the SAW displacements in the new grain. In the parent, the grains are more comparable in size to the excited spot of 140 µm, and have random orientation. Initially SAWs with the same wavelength, but at different frequencies depending on SAW velocity, are generated in different grains within the excitation spot. When a SAW generated in a particular grain propagates into a neighbouring grain with different SAW velocity for that propagation direction, the frequency must stay the same. Hence, in order to match the new SAW velocity, the SAW wavelength must change. The resulting grating with a new wavelength no longer fulfils the diffraction condition for the probe beam, and effectively becomes invisible. Hence the SAWs seem to disappear quickly, but the temperature profile persists. It should be noted that the lifetime of the SAW component in the TGS signal until the SAW leaves the grain is still long enough for it to be picked up in a Fourier transform of the trace. From the traces in Fig. 7(c), we observe six SAW oscillations in the TGS trace from the parent material. With the TGS wavelength of 2.758 μm, this predicts a grain size of 16.5 μm, which is close to the 10-15 μm grain size seen in EBSD (See Fig. 7(b)).

## Implanted High Entropy Alloy

TGS was also used to map a high entropy alloy (HEA) partly implanted with Vanadium ions. An optical micrograph of the sample (recorded using an ALICONA Infinitefocus G5 microscope) is shown in Fig. 8 (a). The alloy consisted of 32.5% Ti, 15% V, 15% Zr, 16.5% Hf and 20% Ta. The sample was produced by arc melting under an Argon atmosphere and then annealed at 1400 °C for 24 hrs under a vacuum of ~0.03 mbar. The sample was ground to 4000 grit and polished with colloidal silica. Diamond paste polishing caused surface pitting too rough for EBSD measurements, and therefore was omitted. The implantations were carried out using 2 MeV $V^+$ ions at 500 °C up to a dose of 3 displacements per atom (dpa) on part of the sample demarcated by a circular scratch. The microstructure was determined to be BCC with rocksalt structured precipitates due to small amounts of carbon and nitrogen infusion from the furnace during fabrication.

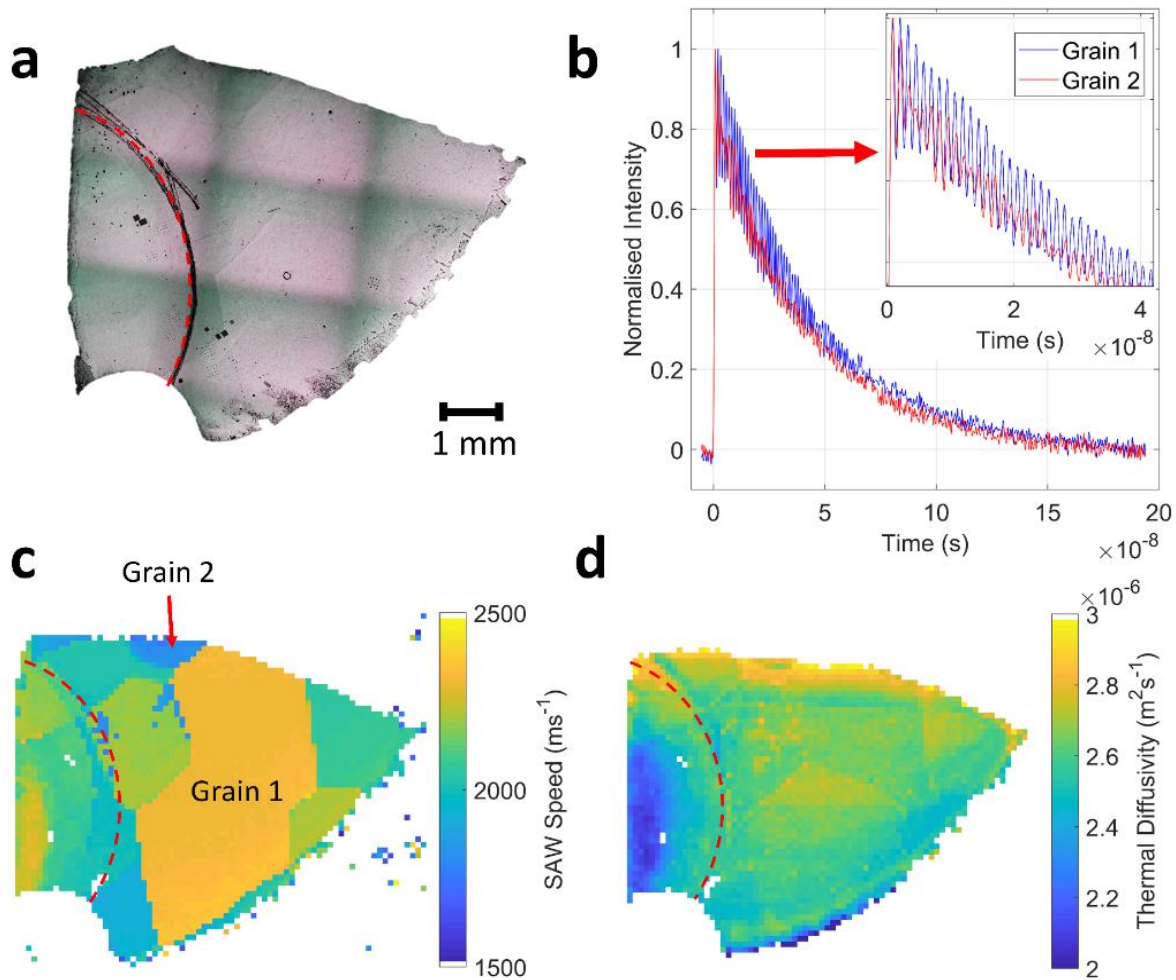

**Figure 8:** Implanted high entropy alloy. Optical micrograph (a), sample TGS traces (b), TGS SAW speed map (c) and TGS thermal diffusivity map (d). The implanted region is indicated by the red dashed line. The same scale bar applies to the micrograph and the TGS maps.

Although the sample surface was too rough for EBSD, the SAW and thermal diffusivity maps, Fig. 8(c) and (d), could be measured with relative ease. Fig. 8(b) shows sample TGS traces for two different grains. A closer look at the traces reveals the variation in SAW frequency.

The scans were carried out using a step size of 200 µm. It is evident from the TGS maps and from the optical micrograph in Fig. 8(a), that the setup could readily map the entire sample, despite the relatively poor surface finish. The SAW speed map clearly shows differences between specific grains, indicating that this HEA must be elastically anisotropic. The boundaries between the regions of different SAW speed closely match the grain boundaries visible in the optical micrograph. The SAW map effectively provides a grain map for elastically anisotropic samples. It also points at the possibility of combining such maps with modelling predictions of SAW speeds as a function of orientation to produce large scale TGS crystal orientation maps [19], [44].

The thermal diffusivity varies little between different grains (Fig. 8(d)). This is expected since cubic crystals have isotropic thermal conductivity [45]. Interestingly the implanted region shows a clear reduction in thermal diffusivity. Metals and metallic alloys are known to be quite susceptible to thermal diffusivity degradation as a result of implantation [15], [46], [47]. This is due to the fact that the predominant thermal energy carries in metals/alloys are electrons, which can be readily scattered by irradiation induced defects. In the SAW map (Fig. 8(c)) an increase in the SAW speed in the irradiated region of the HEA is noticeable. This behaviour is opposite to the reduction in SAW velocity observed for the irradiated regions in the deuterium implanted tungsten sample (See Fig. 5(d)). Unlike the thermal diffusivity, the SAW velocity can increase due to various factors such as stiffening, or decrease due to softening/swelling of the material. The dominant effect depends on the type of material, alloying content, crystal structure, exposure conditions and temperature [9], [48]–[50].

## As Cast Equiatomic TiZrHf Alloy

As a final example we considered another equiatomic alloy composed of TiZrHf, on which EBSD was not possible, which was mapped using the TGS setup. The HEA was produced by arc melting of elemental powders as a 10 g ingot under a vacuum of ~0.03 mbar. The sample was ground to 4000 grit and polished with colloidal silica. Further polishing of the sample was not carried out since diamond paste polishing resulted in surface pitting. EBSD measurements were not possible due to the surface roughness. However, the entire surface area of the sample was mapped using the TGS setup, with a spatial resolution of 200 μm and probing depth of ~1 μm. An optical micrograph of the sample, captured using an ALICONA Infinitefocus G5, is shown in Fig. 9 (a).

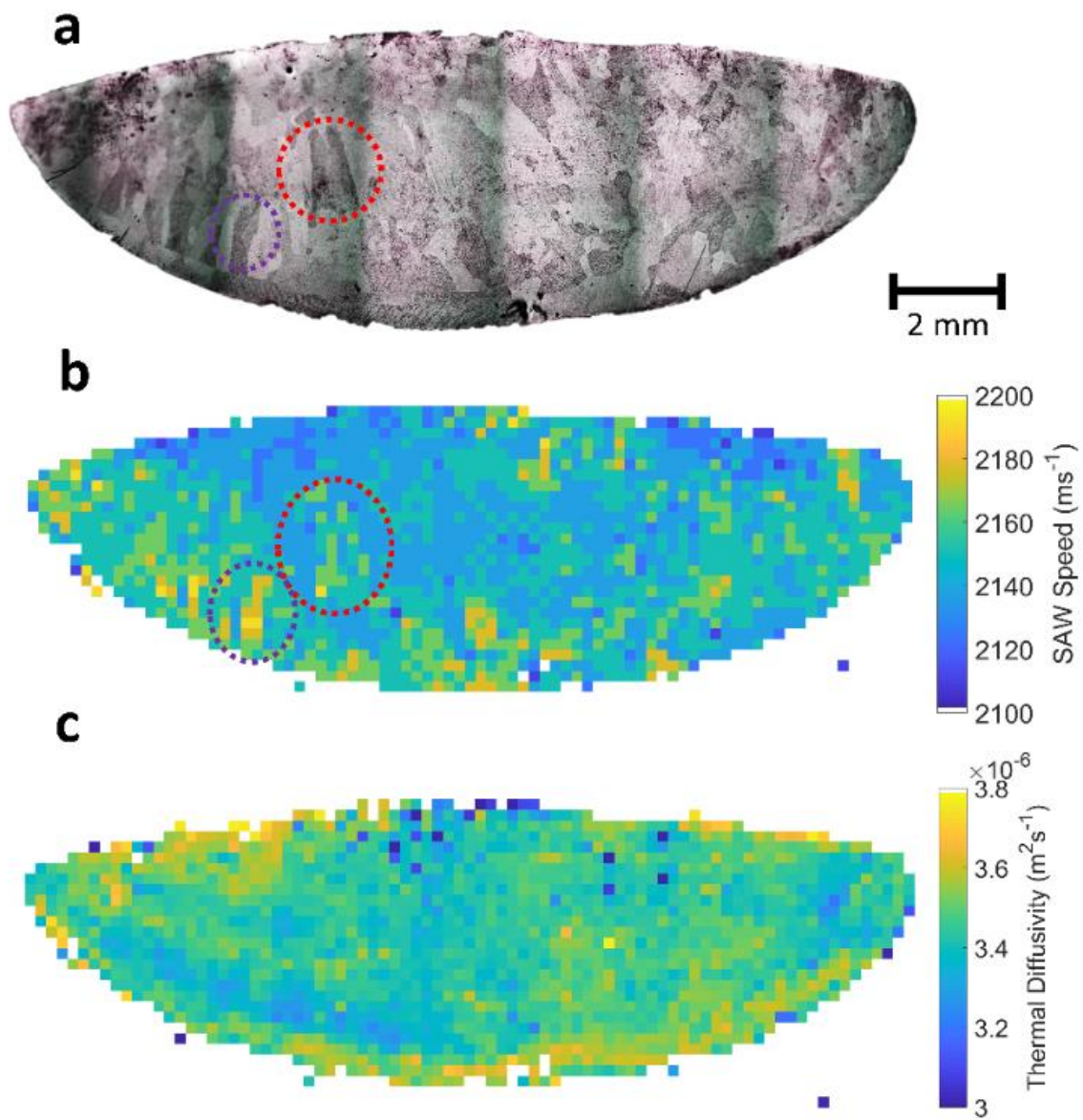

**Figure 9:** TiZrHf equiatomic alloy. Optical micrograph showing different grains (a), SAW speed TGS map (b) and thermal diffusivity TGS map (c). Dashed regions indicate examples of identified grains. The same scale bar applies to the micrographs and the TGS maps.

Fairly large grains are clearly visible in the optical micrograph (Fig. 9(a)). These grains can also be identified in the SAW speed map in Fig. 9(b). This is also due to the elastic anisotropy present, as is the case with most HEA's [51]. The elastic anisotropy evident here is less than that of the HEA in Fig. 8. The darker grains in the optical micrograph can be seen to correspond to regions of higher SAW speed and vice versa. Slight variations in thermal diffusivity are evident across the sample, as expected, due to the HCP crystalline structure. Also evident are lateral bands of higher and lower thermal diffusivity, which may be a cooling rate effect, though this requires further investigation in future.

## Conclusions

We have extended the TGS method to allow the automated measurement of large SAW speed and thermal diffusivity maps. This implementation of TGS as a microscopy tool offers a number of benefits that are successfully demonstrated.

1. Automated large area maps are made across welds, and large scans of implanted areas provide a more comprehensive evaluation of the sample condition. These scans are able to decipher the actual implantation profile, grain orientations, macrozones and pick up anomalies that spot measurements and line scans would fail to show. They also picked up the effects of other measurements, such as NRA, on the sample.
2. In elastically anisotropic samples, quite accurate large area grain maps are can be measured.
3. The high signal-to-noise ratio and automation of scanning and data capture provides many more points temporally and spatially, which results in significantly better statistics than previous TGS, thermal diffusivity or SAW studies.
4. The modified setup is more robust to misalignment as seen with the relatively rough HEA's measured. This is also important for samples with surface slope errors, as well as samples that evolve, e.g. during in-situ heating or deformation.
5. Using the online microscope targeted measurements of specific regions of interest can be made. This allows us to pick out implanted/unimplanted or weld/parent regions in samples for spot measurements as well as to set the boundaries for 2D map measurements. For samples where scans up to an edge are not possible, the microscope allows us to pick a specific point of interest that has been located by another technique, EBSD perhaps, and allows us to directly correlate measurements.

## Acknowledgements

We acknowledge funding from the European Research Council (ERC) under the European Union's Horizon 2020 research and innovation programme (grant agreement No. 714697). The views and opinions expressed herein do not necessarily reflect those of the European Commission.

## References

[1] R. J. Hemley, G. W. Crabtree, and M. V. Buchanan, "Materials in extreme environments," *Phys. Today*, vol. 62, no. 11, pp. 32–37, Nov. 2009.

[2] T. Hirai *et al.*, "ITER divertor materials and manufacturing challenges," *Fusion Eng. Des.*, vol. 125, pp. 250–255, Dec. 2017.

[3] F. Hofmann, M. P. Short, and C. A. Dennett, "Transient grating spectroscopy: An ultrarapid, nondestructive materials evaluation technique," *MRS Bull.*, vol. 44, no. 05, pp. 392–402, May 2019.

[4] O. W. Kading, H. Skurk, A. A. Maznev, and E. Matthias, "Transient Thermal Gratings at Surfaces for Thermal Characterization of Bulk Materials and Thin-Films," *Appl. Phys. a-Materials Sci. Process.*, vol. 61, no. 3, pp. 253–261, 1995.

[5] A. A. Maznev, K. A. Nelson, and J. A. Rogers, "Optical heterodyne detection of laser-induced gratings," *Opt. Lett.*, vol. 23, no. 16, p. 1319, 1998.

[6] J. A. Johnson *et al.*, "Phase-controlled, heterodyne laser-induced transient grating measurements of thermal transport properties in opaque material," *J. Appl. Phys.*, vol. 111, no. 2, 2012.

[7] C. A. Dennett and M. P. Short, "Time-resolved, dual heterodyne phase collection transient grating spectroscopy," *Appl. Phys. Lett.*, vol. 110, no. 21, p. 211106, May 2017.

[8] F. Hofmann *et al.*, "Lattice Swelling and Modulus Change in a Helium - -Implanted Tungsten Alloy : X- - ray Micro- - Diffraction , Surface Acoustic Wave Measurements , and Multiscale modelling," no. 14, 2014.

[9] C. A. Dennett, K. P. So, A. Kushima, D. L. Buller, K. Hattar, and M. P. Short, "Detecting self-ion irradiation-induced void swelling in pure copper using transient grating spectroscopy," *Acta Mater.*, vol. 145, pp. 496–503, Feb. 2018.

[10] J. Goossens *et al.*, "Surface acoustic wave depth profiling of a functionally graded material," *J. Appl. Phys.*, vol. 102, no. 5, 2007.

[11] P.-C. Ostiguy, N. Quaegebeur, and P. Masson, "Non-destructive evaluation of coating thickness using guided waves," *NDT E Int.*, vol. 76, pp. 17–25, Dec. 2015.

[12] M. Brown *et al.*, "Destructive and non-destructive testing methods for characterization and detection of machining-induced white layer: A review paper," *CIRP J. Manuf. Sci. Technol.*, vol. 23, pp. 39–53, Nov. 2018.

[13] F. Hofmann, D. R. Mason, J. K. Eliason, A. A. Maznev, K. A. Nelson, and S. L. Dudarev, "Non-Contact Measurement of Thermal Diffusivity in Ion-Implanted Nuclear Materials," *Sci. Rep.*, vol. 5, pp. 1–7, 2015.

[14] S. E. Ferry, C. A. Dennett, K. B. Woller, and M. P. Short, "Inferring radiation-induced microstructural evolution in single-crystal niobium through changes in thermal transport," *J. Nucl. Mater.*, Jun. 2019.

[15] A. Reza, H. Yu, K. Mizohata, and F. Hofmann, "Thermal diffusivity degradation and point defect density in self-ion implanted tungsten," *arXiv:1909.13612*, Sep. 2019.

[16] A. Reza, Y. Zayachuk, H. Yu, and F. Hofmann, "Transient grating spectroscopy of thermal diffusivity degradation in deuterium implanted tungsten," *Scr. Mater.*, vol. 174, pp. 6–10, Jan. 2020.

[17] J. A. Rogers, A. A. Maznev, M. J. Banet, and K. A. Nelson, "Optical generation and characterization of acoustic waves in thin filsm: Fundamentals and Applications," *Annu. Rev. Mater. Sci.*, vol. 30, no. 1, pp. 117–157, 2000.

[18] R. J. Smith, M. Hirsch, R. Patel, W. Li, A. T. Clare, and S. D. Sharples, "Spatially resolved acoustic spectroscopy for selective laser melting," *J. Mater. Process. Technol.*, vol. 236, pp. 93–102, Oct. 2016.

[19] R. J. Smith, W. Li, J. Coulson, M. Clark, M. G. Somekh, and S. D. Sharples, "Spatially resolved acoustic spectroscopy for rapid imaging of material microstructure and grain orientation," *Meas. Sci. Technol.*, vol. 25, no. 5, 2014.

[20] J. C. Zhao, X. Zheng, and D. G. Cahill, "Thermal conductivity mapping of the Ni-Al system and the beta-NiAl phase in the Ni-Al-Cr system," *Scr. Mater.*, vol. 66, no. 11, pp. 935–938, Jun. 2012.

[21] S. V. Biryukov, Y. V. Gulyaev, V. V. Krylov, and V. P. Plessky, "Basic Types of Surface Acoustic Waves in Solids," Springer, Berlin, Heidelberg, 1995, pp. 1–17.

[22] C. A. Dennett and M. P. Short, "Thermal diffusivity determination using heterodyne phase insensitive transient grating spectroscopy," *J. Appl. Phys.*, vol. 123, 2018.

[23] J. A. Rogers, M. Fuchs, M. J. Banet, J. B. Hanselman, R. Logan, and K. A. Nelson, "Optical system for rapid materials characterization with the transient grating technique: Application to nondestructive evaluation of thin films used in microelectronics," *Appl. Phys. Lett.*, vol. 71, no. 2, pp. 225–227, Jul. 1997.

[24] Y. S. Touloukian, R. W. Powell, C. Y. Ho, and M. C. Nicolaou, *Thermophysical Properties of Matter-The TPRC Data Series. Volume 10. Thermal Diffusivity*. 1974.

[25] M. Fujitsuka, B. Tsuchiya, I. Mutoh, T. Tanabe, and T. Shikama, "Effect of neutron irradiation on thermal diffusivity of tungsten–rhenium alloys," *J. Nucl. Mater.*, vol. 283–287, pp. 1148–1151, Dec. 2000.

[26] M. Fukuda, A. Hasegawa, and S. Nogami, "Thermal properties of pure tungsten and its alloys for fusion applications," *Fusion Eng. Des.*, vol. 132, pp. 1–6, Jul. 2018.

[27] MathWorks, "MATLAB and Simulink Release 2017b." Natick MA, 2017.

[28] R. A. Duncan *et al.*, "Increase in elastic anisotropy of single crystal tungsten upon He-ion implantation measured with laser-generated surface acoustic waves," *Appl. Phys. Lett.*, vol. 109, no. 15, p. 151906, Oct. 2016.

[29] Y. S. Touloukian, R. W. Powell, C. Y. Ho, and M. C. Nicolaou, *Thermophysical Properties of Matter-The TPRC Data Series. Volume 10. Thermal Diffusivity*. Lafayette In: Thermophysical and Electronic Properties Information Analysis Center, 1974.

[30] B. Abad, D.-A. Borca-Tasciuc, and M. S. Martin-Gonzalez, "Non-contact methods for thermal properties measurement," *Renew. Sustain. Energy Rev.*, vol. 76, pp. 1348–1370, Sep. 2017.

[31] Y. Zayachuk, A. Manhard, M. H. J. 't Hoen, W. Jacob, P. A. Zeijlmans van Emmichoven, and G. van Oost, "Depth profiling of the modification induced by high-flux deuterium plasma in tungsten and tungsten–tantalum alloys," *Nucl. Fusion*, vol. 54, no. 12, Dec. 2014.

[32] Y. Zayachuk, I. Tanyeli, S. Van Boxel, K. Bystrov, T. W. Morgan, and S. G. Roberts, "Combined effects of crystallography, heat treatment and surface polishing on blistering in tungsten exposed to high-flux deuterium plasma," *Nucl. Fusion*, vol. 56, no. 8, Aug. 2016.

[33] W. Wang, J. Roth, S. Lindig, and C. . Wu, "Blister formation of tungsten due to ion bombardment," *J. Nucl. Mater.*, vol. 299, no. 2, pp. 124–131, Nov. 2001.

[34] T. Tanabe, C. Eamchotchawalit, C. Busabok, S. Taweethavorn, M. Fujitsuka, and T. Shikama, "Temperature dependence of thermal conductivity in W and W–Re alloys from 300 to 1000 K," *Mater. Lett.*, vol. 57, no. 19, pp. 2950–2953, Jun. 2003.

[35] H.-W. Becker and D. Rogalla, "Nuclear Reaction Analysis," in *Neutron Scattering and Other Nuclear Techniques for Hydrogen in Materials. Neutron Scattering Applications and Techniques.*, Springer, Cham, 2016, pp. 315–336.

[36] V. K. Alimov, M. Mayer, and J. Roth, "Differential cross-section of the D(3He, p)4He nuclear reaction and depth profiling of deuterium up to large depths," *Nucl. Instruments Methods Phys. Res. Sect. B Beam Interact. with Mater. Atoms*, vol. 234, no. 3, pp. 169–175, Jun. 2005.

[37] A. E. Sand, K. Nordlund, and S. L. Dudarev, "Radiation damage production in massive cascades initiated by fusion neutrons in tungsten," *J. Nucl. Mater.*, vol. 455, no. 1–3, pp. 207–211, Dec. 2014.

[38] J. F. Ziegler, M. D. Ziegler, and J. P. Biersack, "SRIM – The stopping and range of ions in matter (2010)," *Nucl. Instruments Methods Phys. Res. Sect. B Beam Interact. with Mater. Atoms*, vol. 268, no. 11–12, pp. 1818–1823, Jun. 2010.

[39] R. S. Mishra and Z. Y. Ma, "Friction stir welding and processing," *Mater. Sci. Eng. R Reports*, vol. 50, no. 1–2, pp. 1–78, Aug. 2005.

[40] S. Bertrand, D. Shahriari, M. Jahazi, and H. Champliaud, "Linear friction welding process simulation of Ti-6Al-4V alloy: a heat transfer analysis of the conditioning phase," *Procedia Manuf.*, vol. 15, pp. 1382–1390, Jan. 2018.

[41] J. L. W. Warwick, N. G. Jones, I. Bantounas, M. Preuss, and D. Dye, "In situ observation of texture and microstructure evolution during rolling and globularization of Ti-6Al-4V," *Acta Mater.*, vol. 61, no. 5, pp. 1603–1615, Mar. 2013.

[42] T. Ben Britton, S. Birosca, M. Preuss, and A. J. Wilkinson, "Electron backscatter diffraction study of dislocation content of a macrozone in hot-rolled Ti-6Al-4V alloy," *Scr. Mater.*, vol. 62, no. 9, pp. 639–642, May 2010.22

[43] E. Wielewski, C. R. Siviour, and N. Petrinic, "On the correlation between macrozones and twinning in Ti-6Al-4V at very high strain rates," *Scr. Mater.*, vol. 67, no. 3, pp. 229–232, Aug. 2012.

[44] W. Li, S. D. Sharples, R. J. Smith, M. Clark, and M. G. Somekh, "Determination of crystallographic orientation of large grain metals with surface acoustic waves," *J. Acoust. Soc. Am.*, vol. 132, no. 2, pp. 738–745, Aug. 2012.

[45] N. W. Ashcroft and N. D. Mermin, *Solid state physics*. Holt, Rinehart and Winston, 1976.

[46] R. K. Williams, R. K. Nanstad, R. S. Graves, and R. G. Berggren, "Irradiation effects on thermal conductivity of a light-water reactor pressure vessel steel," *J. Nucl. Mater.*, vol. 115, no. 2–3, pp. 211–215, Apr. 1983.

[47] A. T. Peacock *et al.*, "Overview of recent European materials R&D activities related to ITER," *J. Nucl. Mater.*, vol. 329–333, no. 1-3 PART A, pp. 173–177, 2004.

[48] F. A. Garner, M. B. Toloczko, and B. H. Sencer, "Comparison of swelling and irradiation creep behavior of fcc-austenitic and bcc-ferritic/martensitic alloys at high neutron exposure," *J. Nucl. Mater.*, vol. 276, no. 1, pp. 123–142, Jan. 2000.

[49] M. Victoria *et al.*, "Microstructure and associated tensile properties of irradiated fcc and bcc metals," *J. Nucl. Mater.*, vol. 276, no. 1, pp. 114–122, Jan. 2000.

[50] C. D. Hardie, C. A. Williams, S. Xu, and S. G. Roberts, "Effects of irradiation temperature and dose rate on the mechanical properties of self-ion implanted Fe and Fe-Cr alloys," *J. Nucl. Mater.*, vol. 439, no. 1–3, pp. 33–40, 2013.

[51] S. Huang, F. Tian, and L. Vitos, "Elasticity of high-entropy alloys from ab initio theory," *J. Mater. Res.*, vol. 33, no. 19, pp. 2938–2953, Oct. 2018.